# Harvesting magneto-acoustic waves using magnetic two-dimensional chromium telluride (CrTe$_3$)

*Chinmayee Chowde Gowda[1], Alexey Kartsev[2,3,4*], Nishant Tiwari[5], Suman Sarkar[6], Safronov A.A[3] Varun Chaudhary[7*] and Chandra Sekhar Tiwary[1,5*]*

[1]School of Nano Science and Technology, Indian Institute of Technology Kharagpur, West Bengal – 721302, India

[2]Computing Center of the Far Eastern Branch of the Russian Academy of Sciences, 680000 Khabarovsk, Russia

[3]MIREA-Russian Technological University, 119454 Moscow, Russia

[4]Peoples' Friendship University of Russia (RUDN University), 6 Miklukho-Maklaya St, Moscow 117198, Russia

[5]Department of Metallurgical and Materials Engineering, Indian Institute of Technology Kharagpur, West Bengal – 721302, India

[6]Department of Materials Engineering, Indian Institute of Technology Jammu, Jammu and Kashmir – 181221, India

[7]Industrial and Materials Science, Chalmers University of Technology, Gothenburg - 412 96 Sweden

Corresponding Email address: Chandra Sekhar Tiwary (chandra.tiwary@metal.iitkgp.ac.in), Alexey Kartsev (karec1@gmail.com) and Varun Chaudhary (varunc@chalmers.se)

**Abstract:**

A vast majority of electrical devices have integrated magnetic units, which generate constant magnetic fields with noticeable vibrations. The majority of existing nanogenerators acquire energy through friction/mechanical forces and most of these instances overlook acoustic vibrations and magnetic fields. Magnetic two-dimensional (2D) tellurides present a wide range of possibilities for devising a potential flexible energy harvester. We have synthesized two-dimensional chromium telluride (2D CrTe$_3$) which exhibits ferromagnetic (FM) nature with a $T_c$ of ~224 K. The structure exhibits stable high remnant magnetization, making 2D CrTe$_3$ flakes a potential material for harvesting of magneto-acoustic waves at room temperature. A magneto-acoustic nanogenerator (MANG) was fabricated composing of 2D CrTe$_3$ dispersed in a polymer matrix. Basic mechanical stability and sensitivity of the device with change in load conditions were tested. A high surface charge density of 2.919 mC m$^{-2}$ was obtained for the device. The thermal strain created in the lattice structure was examined using *in-situ* Raman spectroscopic measurements. The magnetic anisotropy energy (MAE) responsible for long-range FM ordering was calculated with the help of theoretical modelling. The theoretical calculations also showed opening of electronic bandgap which enhances the flexoelectric effects. The MANG can be a potential energy harvester to synergistically tap into the magneto—acoustic vibrations generated from the frequency changes of a vibrating device such as loudspeakers.

## 1.0 Introduction

Two-dimensional (2D) van der Waals (vdW) materials have been shown to have remarkable transport, magnetic, optical, and topological properties, offering exceptional platform for the fabrication of multifunctional devices. Among them, layered Tellurides [1] find applications in spintronic devices [2,3], tunnel junctions [4] and flexible sensors [5]. A few of these 2D tellurides have been employed for energy harvesting such as $Co_2Te_3$, were studies show bandgap opening resulting in enhanced charge storage capability [6,7]. Another study involving $Gd_2Te_3$ embedded 3D printed structure, the nanosheets were able to harvest water droplets due to enhanced surface charge densities [8]. Flexible nanogenerators fabricated using nickel and manganese based di-telluride show high power densities up to 1.89 mWcm$^{-2}$ and 123 mWcm$^{-2}$, respectively [9,10]. Layered $Bi_2Te_3$ was used for thermoelectric power harvesting [11]. A significant portion of our everyday electronics, wearable health monitoring systems and magnetic bit data storage devices, rely on magnetic materials [12–14]. Magnetic field detection has numerous uses in biomedical and portable electronic equipments [15–17]. The interaction between magnetic and electric fields, in such sophisticated devices is often overlooked.

When reduced to monolayer, magnetic materials with interlayer vdW forces can still support ordered magnetic phases. vdW layered ferromagnets exhibit a lower Curie temperature ($T_c$) as compared to bulk crystal forms. The size tunability affects the magnetic behavior of a two-dimensional (2D) material with increased magnetic moment per atom [18,19]. Despite being non-magnetic in bulk, few materials exhibit magnetic behavior in their lower dimensions due to a change in spin domain ordering. Due to the reduced crystal symmetry, many vdW magnets possess magnetocrystalline anisotropy which can be used to devise flexible energy harvesters. Transition metals such as Cr-Te are complex binary systems with vdW layers which can be exfoliated preserving the magnetic anisotropy at ground state [20]. The Cr-Cr distance can be tuned on application of external strains. Chromium tellurides have been well documented for their unique structure and magnetic properties [21–23]. Most of them are metals undergoing transitions to ferromagnets at higher temperatures. Experimental studies on the monolayer Cr samples have been recently reported for compounds such as $CrBr_3$, $CrCl_3$, $CrTe_2$ and $Cr_2Te_3$ [24–27]. These compounds have a weak FM ordering with low Curie temperature ($T_c$). $CrTe_3$ is a tellurium rich compound which is formed due to peritectic reaction at 480 °C. Bulk $CrTe_3$ is a vdW layered antiferromagnet (AFM) with magnetic moment of Cr atoms of 3 – 3.5 $\mu_B$/Cr, whereas the magnetic moment of Te atoms have a very small value which are usually induced by interaction with neighboring Cr atoms [21].

In this work, mechanically exfoliated chromium telluride ($CrTe_3$) which showed FM ordering near RT with high $T_c$ contradicting existing 2D ferromagnets. We further discuss the magnetic behavior of the 2D material. We quantitatively characterize bulk and 2D $CrTe_3$ samples with diffraction, spectroscopic and micro-analysis. A magnetically responsive magneto-acoustic nanogenerator (MANG) was fabricated with the exfoliated 2D $CrTe_3$. The thin film was responsive to low magnetic fields of 1.32 mT. Theoretical modelling was performed to observe the band changes and ferromagnetic ordering at varied temperature ranges. The strain induced flexoelectric charge generation was observed in the 2D $CrTe_3$ MANG which was due to semiconducting nature of the material along with enhanced magnetic anisotropy energy (MAE). Loudspeaker drivers use electromagnetism to transform the audio signal's AC voltage into a magnetic field that moves the diaphragm, creating vibrations. We have devised a potential energy harvester using the ferromagnetic 2D $CrTe_3$ which synergistically combines magnetic moments affecting the spins in the 2D material and acoustic waves (vibrations at varied frequencies) produced form a loudspeaker.

## 2.0 Materials and methods

### 2.1 Experimental details

Stoichiometric amounts of cut tellurium (Te) and chromium (Cr) pieces (99 wt.%) were used as starting materials. Due to the significant difference in the melting points of Cr and Te, both elements are sealed in a quartz tube before melting, with Ar backfilled to stop Te from evaporating. The alloy was inducted melted at 1350 °C in an argon (Ar) atmosphere, and it was held at that temperature for two hours. The samples were subjected to a 75 - 80 hour heat treatment at 900 °C to ensure homogenization.

Isopropyl alcohol (IPA) was used as the solvent in a liquid phase exfoliation procedure to produce the exfoliated sample. The ratio of sample to solvent was maintained at 10 mg/100 ml. A 30 kHz probe sonicator with a 10 sec pulse rate was employed for a duration of 10 to 12 hours. After allowing the sample solution to rest for a full day, the supernatant was gathered and utilized for additional characterization and device construction.

### 2.2 Theoretical Modeling

Structural relaxation procedure was carried out within the framework of the density functional theory (DFT) and projector augmented wave (PAW) method based on the VASP *ab initio* package. The initial *cif*-file of the $CrTe_3$ volumetric sample was taken out from the Material Project database. The following numerical parameters and approaches were employed: 450 eV cutoff energy for the plane-wave-basis, $10^{-8}$ eV/$10^{-7}$ eV/Å energy/force convergence threshold, 8×8×8 k-points grid and the generalized gradient approximation (GGA) with Perdew-Burke-Ernzerhof

(PBE) functional. To properly account the correlation effects the Hubbard correction U=2.65 were applied to Cr *d*-orbitals within the Dudarev formalism. At the starting point of the self-consistent calculations the initial magnetic moment of Cr was chosen to be 3.0 $\mu_B$ and the Te magnetic moment was neglected. In the case of the $CrTe_3$ monolayered structure to avoid interlayer interaction the pseudo vacuum 20 Å was chose along with 1×8×8 k-points grid. The corresponding volumetric and polyhedral models of the bulk $CrTe_3$ are showed on the **Fig. S1(a) and S1(b)** respectively. The non-collinear and spin-orbit coupling (SOC) formalisms were employed to calculate magnetocrystalline anisotropy energy (MAE).

### 2.3 Characterization

An X-ray diffractometer (Bruker D8 Advance) was used to measure X-ray diffraction patterns and to provide information on crystalline phases. At 40 kV voltage and 40 mA current, the Cu-K$\alpha$ source's wavelength was ($\lambda$) 1.5406 Å. High resolution transmission electron microscopy (HRTEM) imaging was carried out in a JOEL F30 FEI unit operating at 300 kV. Samples that had been exfoliated were dispersed in ethanol and sonicated for ten minutes. Subsequently, the dispersed sample was drop-cast in one or two drops onto the copper grid. The copper grid used in the TEM was coated with carbon (holey carbon). The specification is 500 square mesh. Agilent Technologies Model No. 5500 was used to perform atomic force microscopy (AFM). Using a WITec Raman spectrometer (WITec, UHTS 300 VIS, Germany) at room temperature (RT) and a laser excitation wavelength of 532 nm, a Raman spectroscopy analysis was performed. Using Al–K$\alpha$ radiation ($\lambda$=1486.71 eV) as the source. With the aid of an analytical UV-visible spectrophotometer, absorbance and band gap measurements are performed. An XPS ThermoFisher Scientific Nexsa was utilized to examine the oxidation states and surface composition of the materials. Magnetic measurements were performed using a Quantum Design MPMS SQUID VSM

EverCool system with an operating temperature range from 1.8 K to 1000 K. Voltage measurements were performed using DSO Textronix 1072B. The electrical measurements were performed using precision LCR meter SM6026 (Scientific). Litesizer 500 (Anton paar) was used for Zeta potential measurement. Squidstat plus potentiostat was used for galvonostatic electrochemical impedance spectroscopy (EIS) measurements and Zahner analysis software was used for fitting of EIS data.

**2.4 Fabrication of magneto-acoustic nanogenerator (MANG)**

The exfoliated powder gathered from the supernatant of the material is mixed with a soft polymer matrix; polyvinylidene fluoride (PVDF). The ratio of the material to binder is optimized at 1 mg of 2D $CrTe_3$ to 10 ml of PVDF solution to fabricate a thin film. The thin film was molded to a very thin sheet and also in different bend forms where it can be attached on the body parts (e.g., hands). The thickness of the 2D layer was very thin ($< 0.5$ mm). *Kapton* tape was used as counter electrode to fabricate the flexible nanogenerator. Copper tape was used as external contacts at each end. The electrical impulses generated during application of load, frequency and magnetic fields are measured using a digital oscilloscope (DSO) via copper contacts connected at each end of the film.

## 3.0 Results and discussion

### 3.1 Physical characterization

The bulk sample was alloyed using induction melting technique at 1350 °C, the sample phase formation was confirmed from X-ray diffractograms (XRD) and scanning electron microscopic (SEM) with elemental distribution (EDS) analysis as seen in **SI Fig. S2 (a – c)**. The samples were then crushed and exfoliated using probe sonication for a period of 10 - 12 h. The sample was suspended for 48 h, which was centrifuged, dried and collected for further characterization. The exfoliated 2D $CrTe_3$ structure was created based on a visual analysis of its bulk prototype (**Fig. 1a**), where vdW layers are lying in the (100) crystallographic planes with the distance between them ~ 3 Å. Therefore, (100) is expected to be the prioritized plane of exfoliation for the $CrTe_3$. The layers are usually made up of $Cr_4Te_{16}$ units comprising four edge-sharing $CrTe_6$ octahedra. The octahedra share corners with those in neighboring $Cr_4Te_{16}$ units to form the 2D layers [28]. **Fig. 1b inset** shows Cr-centered polyhedral model of the 2D-$CrTe_3$ structure. The experimental confirmation of the highest exfoliated planes was observed from XRD pattern obtained for the exfoliated $CrTe_3$ as seen in **Fig. 1b**. We have ruled out the existence of other phases of chromium tellurides by comparing the experimental data and theoretical simulations. The diffraction peaks were matched with JCPDS card no: 01-074-1949. For further morphological analysis, the sample dispersion was drop casted on a holey carbon grid with 500 mesh. **Fig. 1c** shows HRTEM image of $CrTe_3$ flake with FFT pattern (**Fig. 1c inset**). The FFT shows stacking of monoclinic pattern. The line profile shows ~140 pm corresponding to atomic radii of Te atoms as seen in **Fig. 1d**. A small section of the flake is processed further using inverse FFT in **Fig. 1e**. The fringes are approximately spaced at $d = 0.36$ nm corresponding to (100) plane which was also tabulated form DFT calculations.

X-ray photoelectron spectroscopy was performed to confirm the formation of $CrTe_3$ composition. The exfoliated 2D $CrTe_3$ sample showed peaks for both Cr and Te elements in the XPS analysis. The XPS Te3d and core level spectra, as well as the overlap with the Cr2p core level, show evidence of surface oxidation peaks. Because of the overlap of the Te3d signals with the Cr2p core level, this is regarded as an estimate due to uncertainties in deconvoluting the peaks. Cr2p is depicted in **Fig. 1f**, where satellite peaks at 574 and 584 eV are seen alongside peaks at 577 eV ($2p_{3/2}$) and 586 eV ($2p_{1/2}$) with FWHM of 3.1 and 3.5, respectively. Similar to this, Te3d peaks are seen in **Fig. 1g** at 577.35 eV for $3d_{1/2}$ and overlaps with the $Cr2p_{3/2}$ peak at $3d_{5/2}$. To confirm the thickness of the exfoliated flakes, atomic force microscopy (AFM) is performed in non-contact mode. **Fig. 1h** shows flakes dispersed and the average thickness of the flakes is mapped to be 6 – 7 nm (from the scale). The lateral area of the flakes is also observed (**Fig. 1i**) to be 60 - 75 nm. The pictorial description of the exfoliated sample dispersion is as seen in **Fig. 1i inset**.

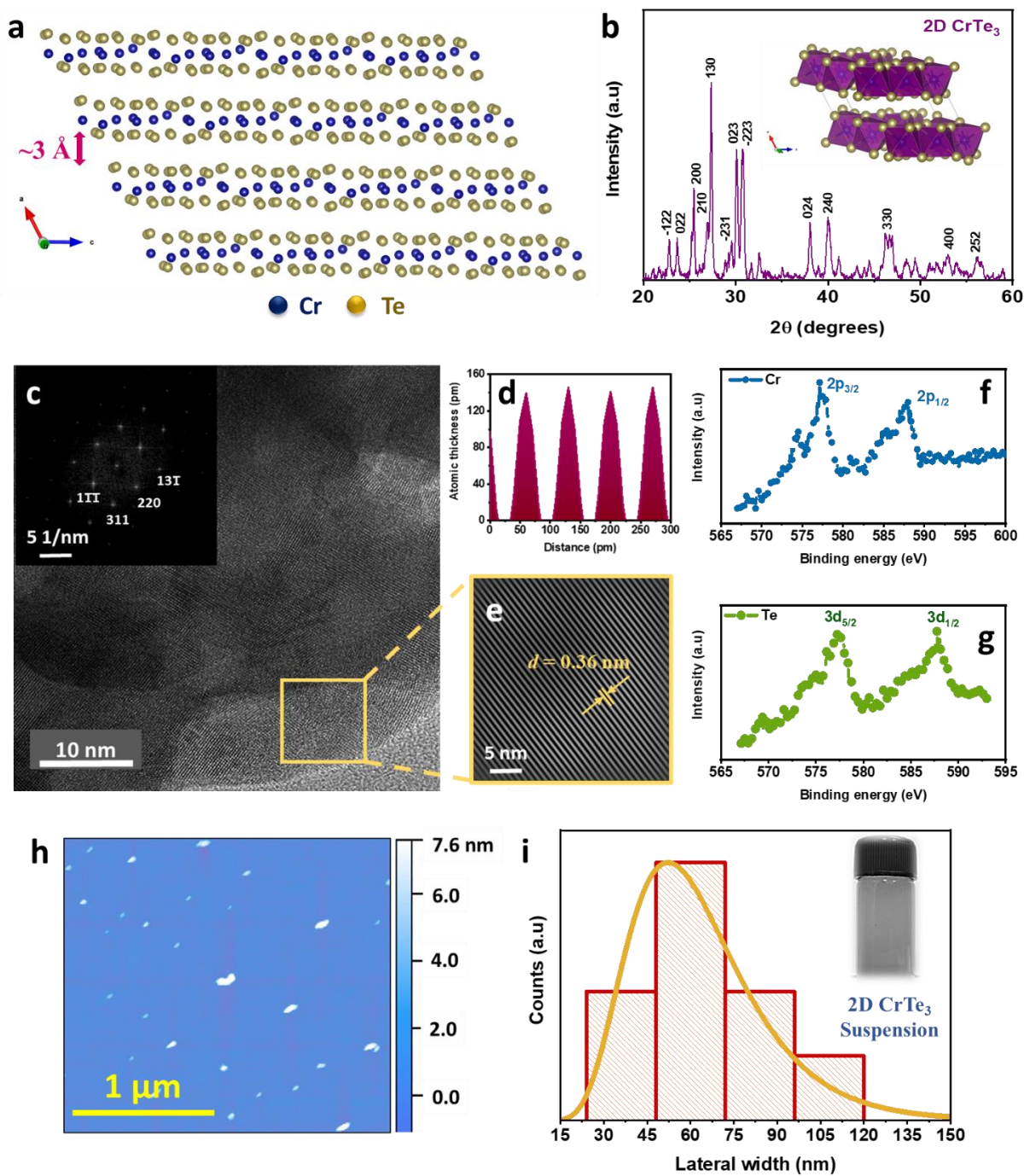

*Fig. 1: Structural and spectroscopic composition of 2D CrTe₃:* (a) CrTe₃ (100)-layered structure, blue and yellow balls indicate Cr and Te atoms, respectively, (b) X-ray diffraction pattern with inset of Cr-centered polyhedral model of the 2D-CrTe₃ structure, (c) HRTEM image of the 2D CrTe₃ flake, (d) inverse FFT derived from the pattern in **(c)**, (e) Atomic radii of Te

*atoms, (f,g) X-ray photoelectron spectroscopy of Cr and Te, respectively, (h) AFM showing flakes with ~ 4nm thickness and (i) Graph showing lateral dimension of the dispersed flakes, inset: Liquid phase exfoliated sample in IPA solvent.*

## 3.2 Theoretical Modelling

The magnetic anisotropy energy (MAE) was calculated for two different *(100)* and *(010)* crystallographic planes. **Fig. 2a** and **2b** represents the $E^{tot}[\cos(\theta)]$ showing the total energy (per formula unit) dependence on the $\theta = \widehat{(\vec{\mu}(Cr), \vec{c})}$ angle cosine, where $\vec{\mu}(Cr)$ is the collinearly allied Cr-magnetic moments direction and $\vec{c}$ is the unit-cell vector. The electronic properties of bulk and monolayered $CrTe_3$ structures were also investigated. The 3D charge density distribution is shown in **Fig. 2c** for the bulk and 2D structure in **Fig. 2d**, correspondingly. The charge density redistribution in the 1L-$CrTe_3$ is observed in comparison to the bulk structure, which correspond to the higher degree of bonding between Te and Cr atoms in the monolayered structure.

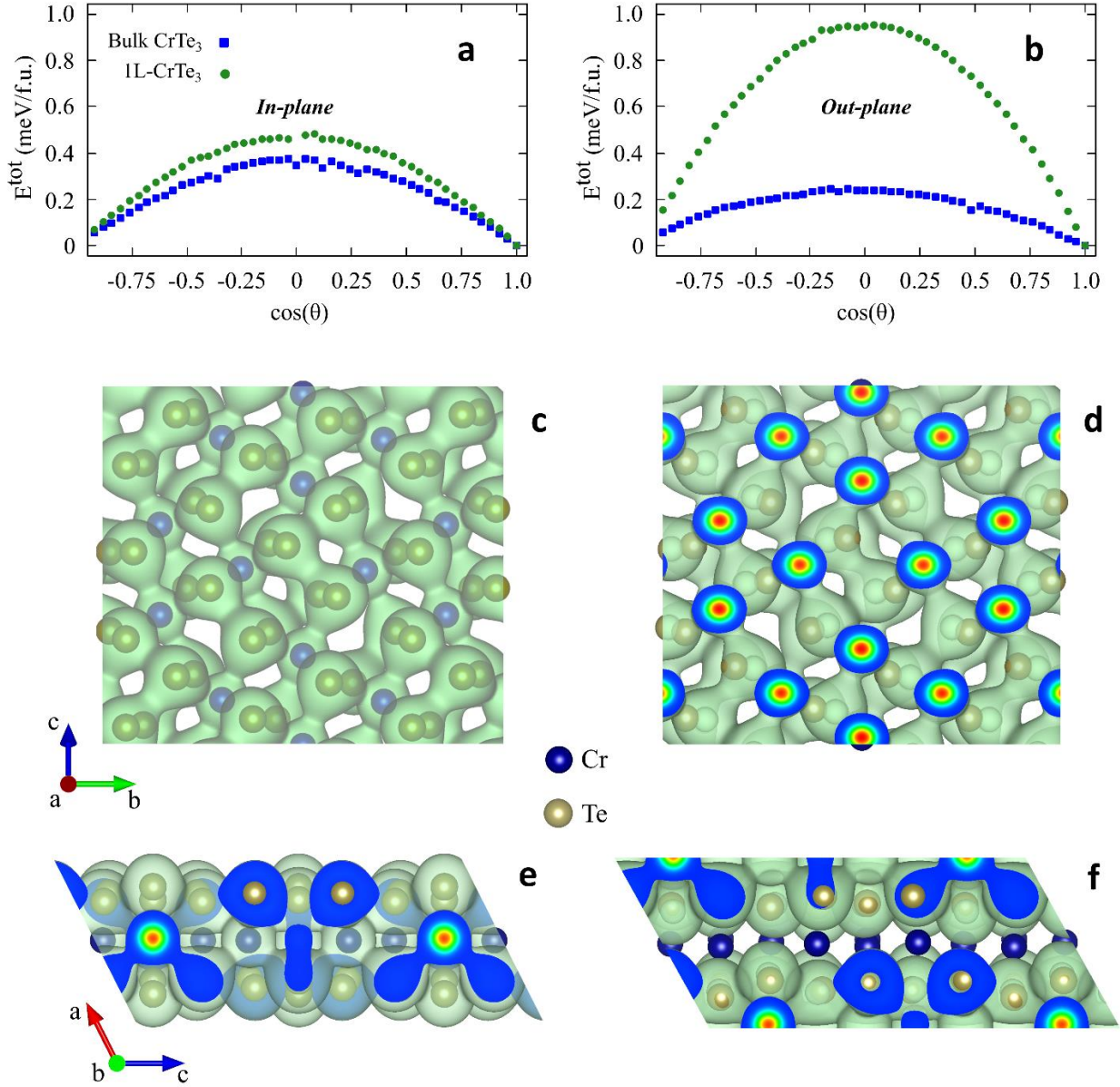

**Fig. 2: Theoretical calculations:** Total energy $E^{tot}[\cos(\theta)]$ as a function of the $\theta = \widehat{(\vec{\mu}(Cr),\vec{c})}$ angle cosine calculated for the bulk and 2D $CrTe_3$ structures in the (100) (a) and (010) planes (b) correspondingly. The reference point is chosen to be equal to the ground state energy value, 3D charge density isosurface (0.04 e/Å³) for the 2D-$CrTe_3$ top- (c) and side-view (e), and top- (d) and side-view (f) for the bulk structure correspondingly. Blue and yellow balls indicate Cr and Te atoms respectively.

### 3.3 Magnetic and optical studies

Superconducting quantum interference device (SQUID) measurements were performed to obtain $M-H$ loops. In **Fig. 3a,** magnetization ($M$-$H$) curves of the bulk and 2D CrTe$_3$ were measured at room temperature with varying the applied magnetic field within the range of −60 and 60 kOe. The 2D CrTe$_3$ was further probed to study its behavior at varied lower temperatures. We measure $M$-$H$ curves, with a step change in temperature of 50 K, ranging from 300 to 10 K (**Fig. 3b**) with an applied field of -20 to 20 kOe. The sample shows highly ferromagnetic (FM) behavior at lower temperatures with broader hysteresis as observed in **Fig. 3b**. The magnetic nature of the material is decided based on the coupling between the magnetic moments which are induced in the lattice structure. Long-range magnetic ordering can be stabilized by magnetic anisotropy with excitation bandgap opening which was observed in the 2D CrTe$_3$ samples. **Fig. 3c** shows mapping of coercivity ($H_c$) and magnetic saturation ($M_s$) increased with decrease in temperature. $H_c$ on the other hand was 4.56 kOe at lower temperature (10 K) and 41.4 Oe at room temperature. Magnetic saturation ($M_s$) of exfoliated CrTe$_3$ was increased from 2.43 (250 K) to 7.87 emu g$^{-1}$ (10 K). Because of the disruption of magnetic ordering at lower temperatures, the $M_s$ value at lower temperatures increases. **Fig. 3c inset** shows Stoner-Wohlfarth value plotting the ratio of magnetic remanence ($M_r$) and $M_s$ of the sample at varied temperatures. The higher the ratio defines the squareness of hysteresis loop formed and its utilization in memory based devices, due to its data storage capacity. $M_r/M_s$ ratio was around 0.0025 at 250 K and in the range of 0.022 to 0.41 (< 200 K), which is the ideal condition for Stoner-Wohlfarth (0.5) is stating randomly oriented uniaxial grains in the atomically thin flakes. The flake orientation is dependent on the angular field switching at which the magnetic fields are exposed on the flakes [29]. The flakes at lower temperatures have multiple magnetic domains.

Temperature dependent magnetic behavior of the samples were carried out for the exfoliated sample. *FC (Field cooled) - ZFC (zero field cooled)* curves were noted in the range of 350 K to 10 K at applied magnetic field of 1000 Oe. Magnetization curves at varied temperatures *(M-T)* is as seen in **Fig. S4**. Curie temperature ($T_c$) was noted to be 224 K which was slightly lower than the existing ML CrTe$_3$ reported values of 231.5 K [21]. Inverse magnetic susceptibility ($\chi^{-1}$) curves were plotted versus temperature (**Fig. 3d**), and we observe that the exfoliated sample follows Curie-Weiss form at high temperatures above $T_c$. Curie–Weiss form is given by $\chi \approx C/(T - \Theta_{CW})$, where $T$ is temperature and $C$ is the Curie constant. This allows extraction of the Curie–Weiss temperature, $\Theta_{CW}$, from a plot of $1/\chi$ versus $T$. Below $T_c$ there is uneven ordering which suggests FM clusters. The low-temperature behavior likely arises from activation of defects acting as donors or acceptors. In order to preserve long-range magnetic ordering in 2D materials, magnetic anisotropy is a prerequisite. The exchange coupling coefficient is positive in this case as the spins tend to be parallel. Due to increase in magnetization above $\theta_{CW}$ (224 K) massive amount of magnons will be generated by the thermal energy, leading to the destruction of long-range order. And below $T_c$ inhibition of thermal excitation was observed for the stabilization of the FM ordering. A stable magnetic range for operation of spintronic application with high remnant magnetization can be established. Many chromium compounds such as CrI$_3$, Cr$_2$Ge$_2$Te$_6$ at lower dimensions have lower $T_c$ (**Fig. 3e**) and as the number of layers decreases the structure loses its remnant magnetization making it antiferromagnetic (AFM) in nature. Various groups have theoretically predicted 2D materials with higher $T_c$ which includes two large groups, TMD Janus structures [30] and Cr compounds [31]. 2D CrTe$_3$ shows higher $T_c$ and $M_r$ which remained stable for thin 2D CrTe$_3$ flakes (~ 4 nm range) near room temperature.

We performed Raman analysis of bare sample (2D $CrTe_3$) on the dispersed sample solution. Evident peaks corresponding to $CrTe_3$ were found with relatively broad signals for 532 nm excitation wavelength at 96.06, 116.49, 128.34 and 201.24 cm$^{-1}$ [32]. We then used the same Raman spectroscopic technique to investigate structural changes at temperatures higher than RT (>298 K). Thermal effects on the thin $CrTe_3$ flakes cause the major Raman peaks to diminish, as shown in **Fig. 3f**. $CrTe_3$'s crystal structure changed as a result of heating effects that led to crystallization. Previous studies have shown that $CrTe_3$ thin film samples crystallize more than thicker ones, as the critical thickness of the materials defines temperature of crystallization [33]. The diffusion at these lower temperatures produces kinetically more stable $CrTe_3$ compounds. In few-layer flakes, the layers initially inter diffuse, resulting in nucleation in the structure, as seen in Raman signatures. The Raman spectral study also aids in determining the stable operating range for device applications.

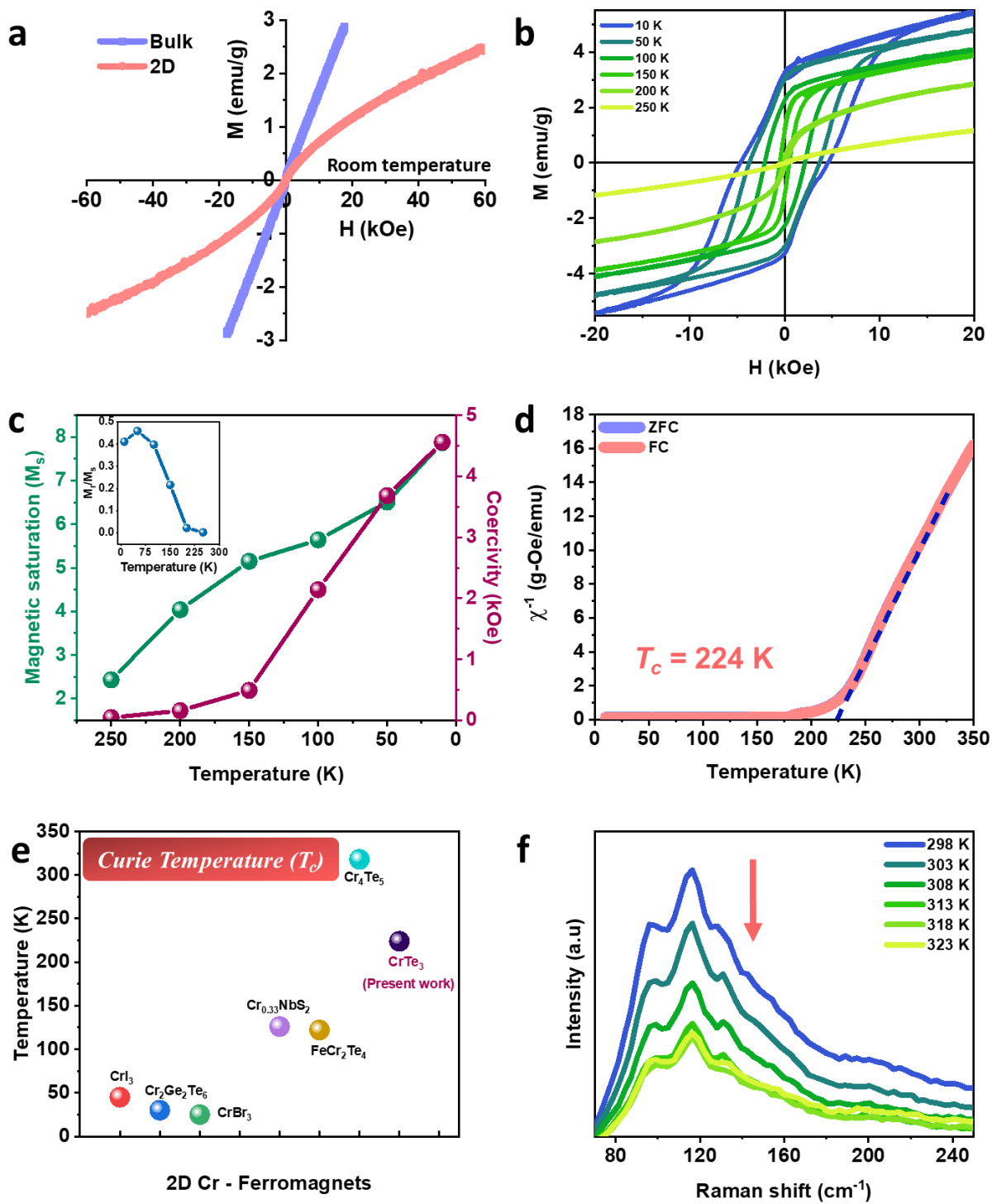

*Fig. 3: Magnetic behavior and optical studies:* (a) M-H loops of bulk and exfoliated CrTe₃ at room temperature (b) M-H of 2D CrTe₃ at varied temperatures, (c) $M_s$ and $H_c$ for 2D CrTe₃ at varied temperatures inset: Stoner- Wohlfarth ratio, (d) Inverse magnetic susceptibility with $T_c$ of

*224 K, (e) T$_c$ for lower dimensional Chromium compounds, and (f) Raman signature of the flakes at higher temperatures.*

## 4.0 Magneto-acoustic Nanogenerator (MANG)

### 4.1 Load based measurements

Nanogenerators (NGs) have a wide range of materials ranging from lead-zirconate-titanate (PZTs) to polymer-based NGs. The mechanical flexibility, toughness, and abrasion resistance of polymer-based energy harvesters are improved, and their biocompatibility makes materials like polyvinylidene fluoride (PVDF) promising for use in environmental friendly applications [15,34–37]. Magnetic nature of a 2D material and its behavioral effects on the device performance is not much explored in these NGs. A limited number of studies demonstrate the optimal harvesters used in electromagnetic fields due to their fixed design and complex mechanism for harvesting kinetic energies [38,39]. Therefore a more interchangeable model with flexible device design is preferred for a wide range of applicability. The NG's design dynamics are different from one another, making flexibility a prerequisite for any such applications. We incorporated 0.5 mg 2D CrTe$_3$ powder in 10 mg/ml concentration of PVDF in acetone. **Fig. 4a** shows the molded thin transparent film with thickness less than 500 microns. The film was magnetically responsive when a Neodymium magnet (N35) of dimensions 20 x 10 x 2 mm with 1.23 T was held nearby as seen in **Fig. 4 (b, c)**. The thin films showed magnetic response. A magneto-acoustic nanogenerator (MANG) was fabricated with the 2D CrTe$_3$ thin film as electrode and *Kapton* as counter electrode with Cu contacts (*Details in **Materials and Methods***). Surface charge density (SCD) was calculated for the device using the formula given in **SI Section 2.2** [35,40]. A high SCD was obtained for the device, the value was calculated to be 1.203 mC m$^{-2}$ for the substrate and 2D

CrTe$_3$. The SCD is inversely proportional to the thickness of the dielectric layer (*d*) as it decreases. After the addition of *Kapton* (1.716 mC m$^{-2}$), the SCD was 2.919 mC m$^{-2}$ for the device. The MANG was tested for its mechanical load bearing abilities at various loads (5, 10, 20, 50, 100, 200 weight cm$^{-2}$) as seen in **Fig. 4d**. The suspended load creates a bending angle in the device which results in voltage generation of 3.4 V$_{p-p}$ at high loads up to 200 weight cm$^{-2}$. This induces a change in capacitance, resulting in an increased bending radius, indicating mechanical sensing capability. Pronounced flexoelectric nature in lower dimensional materials have been well established and is responsible for enhanced voltage outputs [41]. The bend in the device facilitates charge separation as strain gradients arise in very thin films without external strain inducers [42]. In **Fig. 4d inset,** the red line shows near 0 V voltage generated for the *blank* sample (Polymer without 2D CrTe$_3$). The voltage generation due to different loads and angle bends are as depicted in **Fig. 4d**. The voltage-pressure correlation was fitted using a nonlinear curve (Levenberg – Marquardt). More active sites are made available in the 2D CrTe$_3$ for charge conduction as surface oxygen vacancies increase in the 2D material post-exfoliation [34,43–45]. Analog to a magnetic sensor, the response need not be linear in case of magnetic energy harvesters.

We measured the sensitivity of the device at these different load conditions, which resulted in the highest sensitivity of the device being 2039.6 mV kPa$^{-1}$ at 10 g cm$^{-2}$ (**Fig. 4e**) (*Details in SI Section 2.1*). Triboelectric nanogenerators activated by various sources have been used in biomedical applications [46–49]. The MANG can be potentially used as a splint to sense the slightest movement of the joint when attached to the finger (Index) as shown in **Fig. S5 (a, b)**. Splints are usually devised to immobilize and protect the joint injured or for people suffering from rheumatoid arthritis to help in movement. Constant movement can help strengthen the joint while engaging in low impact activities. The bend produces an open circuit voltage (V$_{oc}$) of ~2.7 V as seen in **Fig. 4f**. Positive 1.35 V during downward bend and negative 1.35 V while upward movement of the finger, remains 0 V at no movement (**Fig. 4f inset** shows rest position). Hence the device can be used as a compact mechanical sensor to monitor joints health. The slightest strain induced (angle bend) resulted in considerable voltage output. Under the inhomogeneous deformation in the material, dipole moment is induced via flexoelectric effect. Depending on the polarity of the material the flexoelectric effects can either enhance or suppress piezo response.

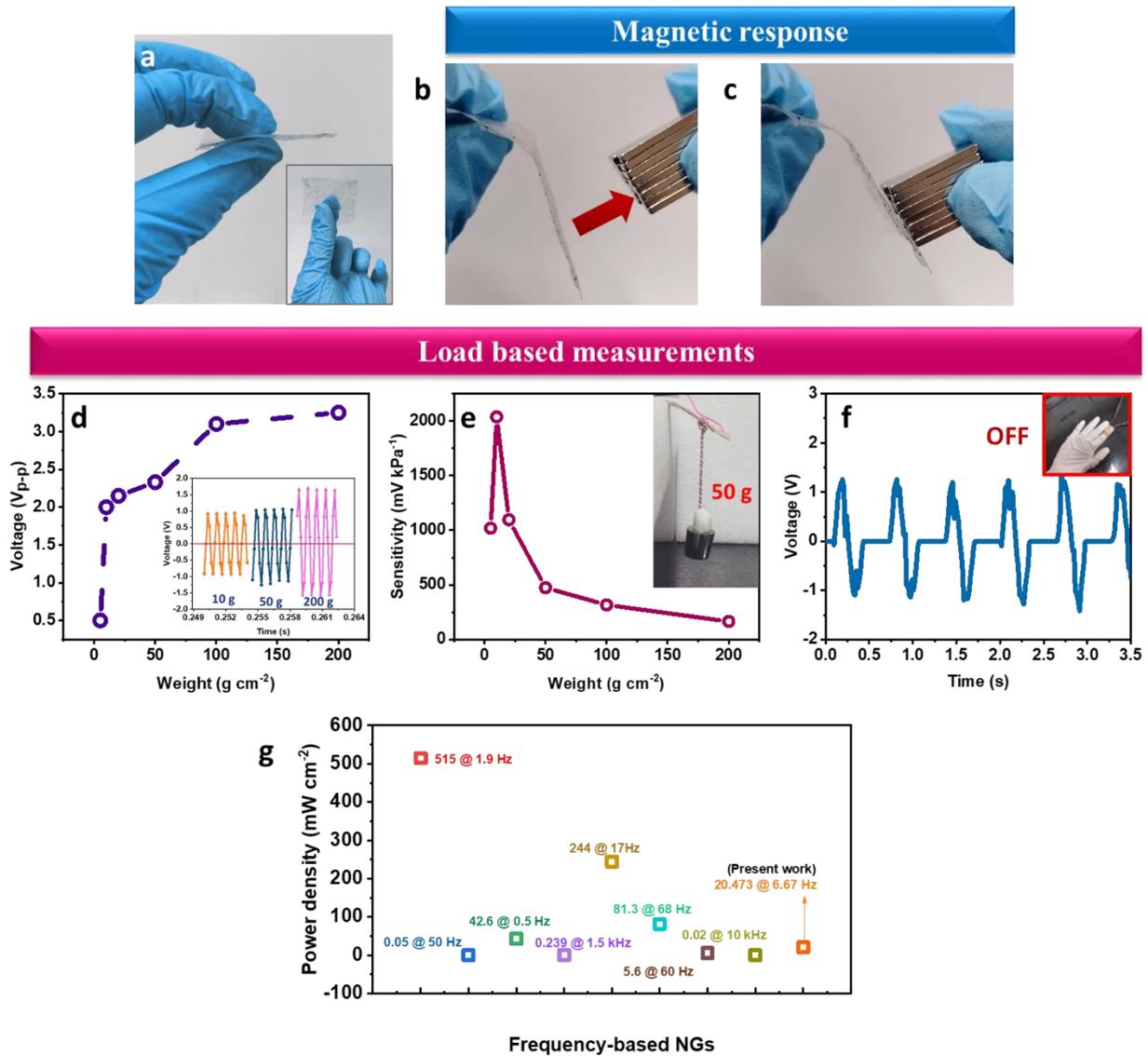

***Fig. 4: Device Measurements:*** *(a) Thin film fabricated from 2D CrTe₃ in polymer matrix, (b,c) The thin film shows magnetic response at 1.32 mT, (d) Voltage output ($V_{p-p}$) of the loads suspended; inset: voltage outputs for loads suspended from the device weighing 10, 20 and 200 g, respectively, (e) Sensitivity of the device inset shows suspended weight of 50 g, (f) The slightest movement in the MANG results in 1.6 V generation inset shows device in rest condition attached to a glove (OFF state) and (g) Comparison of various low-frequency NGs.*

### 4.2 Frequency based measurements

Many NGs have been devised till date to harvest various range of frequency vibrations [40,50]. **Fig. 4g** shows a plot of maximum power density of frequency based energy harvesters at a particular load resistance at various frequency of operation [50–57] as compared to the present work. The 2D CrTe₃ produces a maximum power density of 20 mW cm$^{-2}$ for a low-resonant frequency of 6.67 Hz. Loud speakers are a type of transducer which transfers the electric signal into acoustic signals that is a source of varied low frequencies. Permanent magnets play a major role in determining the speaker's sound quality. Based on the transducer principle, speakers can be categorized as internal or external magnetic based on their magnetic circuit structure, which includes moving-coil, electromagnetic, piezoelectric, and capacitive types. Equipped with electromagnetic driving, the early moving-coil speakers became the most significant vocal style of modern speakers. Moving-type speakers transitioned from electromagnetic driving to permanent driving with the introduction of AlNiCo magnets.

A schematic representation of acoustic vibrations induced by the speaker is shown in **Fig. 5a**. The vibration displacement of the speaker cone has been converted into the vibration velocity and used as a boundary condition for the acoustic wave analysis. Due to electrical energy that is constantly fluctuating, a coil connected to the cone will move forward or backward, changing the air density and producing vibrations as a result. When various frequencies are transmitted to the coil, the coil's energy will interact with the magnetic field of the permanent magnet. This interaction will cause the cone to vibrate further. **Fig. 5b** shows the setup with 2D MANG device attached to the top of the cone which is an extent of the diaphragm of the loudspeaker. The greater the magnetic flux density, the higher the relative power and the higher the sensitivity of the speaker. The permanent magnets inside the loudspeakers can vary from 0.001 tesla (T) all the way up to about 1.5 T. The magnets are simply designed to create vibration that in turn creates sound (acoustic wave). Neodymium magnets were employed as proof masses for generating a magnetic torque under an AC magnetic field; they accelerated the vibration of the diaphragm cone continuously under the effect of the surrounding AC magnetic field. The device attached to a Bluetooth speaker and operating at varied frequencies was also measured at 50 Hz and 500 Hz, respectively in **Supplementary Video 1 and 2**. For most speakers to function as efficient transducers, electromagnetism is a prerequisite. Frequencies were varied in low-frequency range from 1 Hz to 1000 Hz (**Fig. 5c**). $V_{p-p}$ of 6.5 V was obtained for frequency of 1000 Hz and the device saturates thereafter. **Fig. 5d** shows that the measured power output of 2D MANG was 40 µW with currents up to 6.5 µA with varied frequency. Using an LCR unit, electrical measurements such as capacitance and dielectric constant of the 2D MANG device were plotted in **Fig. 5d**. Highest capacitance of 5.73 pF with dielectric constant of 4320 was obtained for the MANG device at 1 Hz resonant frequency.

The operation temperature of the device also plays an important role in the device characteristics. We measured the temperature effects on the 2D MANG as it is made of a very thin polymer matrix (**Fig. 4a**). The operation range of the device was kept from room temperature 25 °C (298 K) till 55 °C (328 K). We observe voltage generation of 0.3 V to 8.1 V when there was a change in temperature from 313 K to 328 K, respectively (**Fig. 5e**). As observed from Raman studies before in **Fig. 3f**, when temperature increased the ordering of phonons stiffened as the vibrations from the cone (loudspeaker) effects the thermal fluctuations in the atoms. Under non-resonant conditions the intensities become weak. Hence, the device remains operable until 45 - 50 °C without changing the crystallinity of the $CrTe_3$ compound.

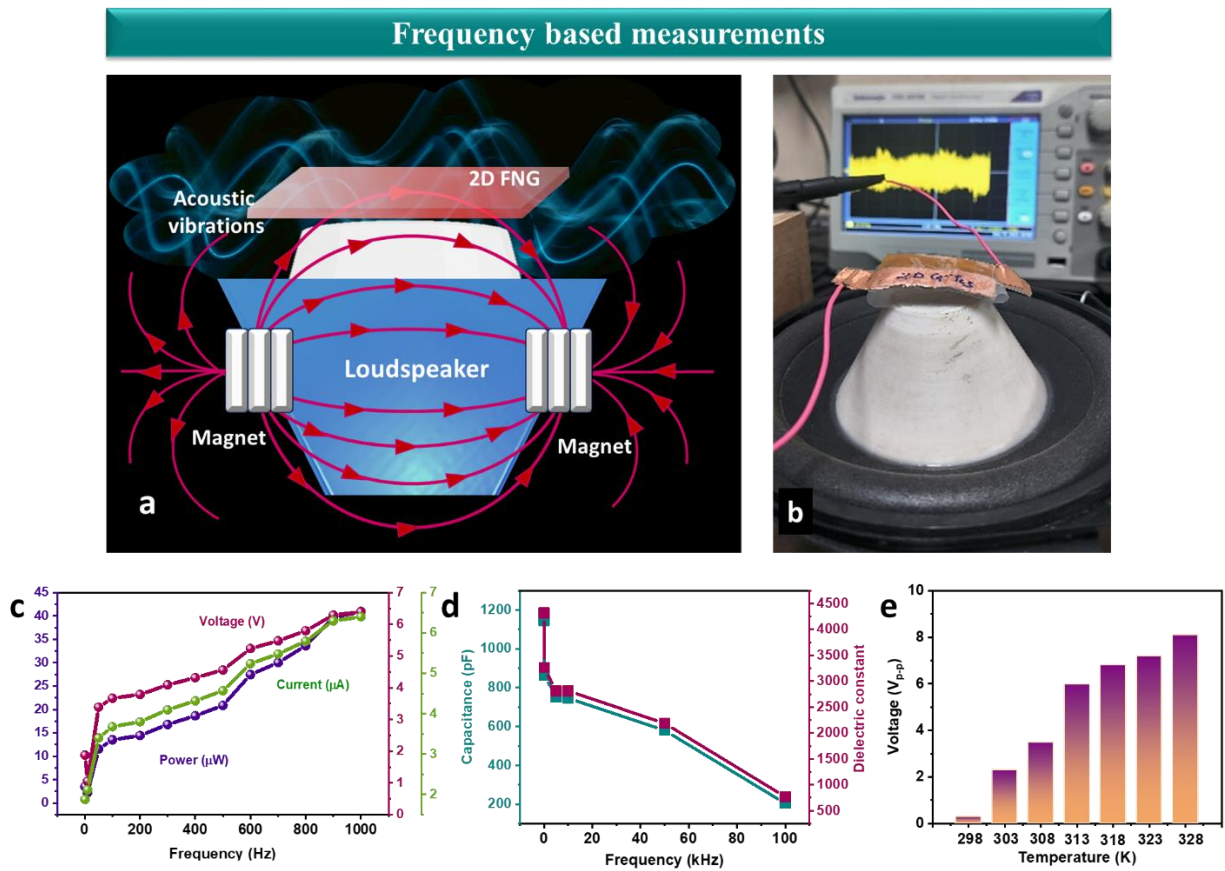

*Fig. 5: Device measurements: Frequency based voltage generation when connected to loudspeaker (a) Schematic representation of charge generation due to magneto-acoustic waves from a loudspeaker. (b) Pictorial representation of MANG device operation, (b) Voltage, Power and Current mapped with variation of frequency, (d) Capacitance and dielectric measurements and (e) Temperature effects on voltage generation.*

**4.3 Device mechanism**

To further explore the role of lattice structure in magnetic anisotropy, the contribution of vacancies and defects effecting the magnetic fields were also studied as shown in **SI Section 3, Fig. S6**. We investigate the effect of point defects on the magnetic anisotropy of a 2D structure. To do this, we consider two cases – remove one Cr/Te atom from the 2D structure. In **Fig. S6** the magnetic anisotropy of the two-dimensional $CrTe_3$ structure is compared with 2D structures whose cells have a point defect Cr **(a)** point defect Te **(b)**. As observed from the **Fig. S6** the Cr atom defect does not lead to any changes in the magnetic anisotropy, while the Te atom defect leads to a fivefold decrease in anisotropy, which allows us to conclude that the non-magnetic environment Te of the Cr magnetic sublattice in a two-dimensional $CrTe_3$ material has a substantial effect on the magnetic anisotropy.

From the experimental results we observe that the remnant magnetization of samples can persist for quite a long time even at RT. The RT is slightly above the critical temperature of 1L-$CrTe_3$ (~224 K), and therefore the FM order can be easily repaired by an external magnetic field. We have studied the impact of external magnetic field in the $CrTe_3$ electronic structure by computing and comparing the density of states (DOS) both for spin polarized (FM) and non-spin polarized (NM) cases including SOC effects. NM solution corresponds to the paramagnetic state of $CrTe_3$ with disordered spin structure. While FM mimics the 1L-$CrTe_3$ with a restored long-range FM order in the external magnetic field slightly above the critical temperature. From the **Fig. 6** the band gap opening (~1 eV) in the electronic structure occurs along with magnetic subsystem ordering due to the $p$ and $d$-hybridization. Therefore, causing stronger piezoelectric effects due to the FM 1L-$CrTe_3$ semiconductor nature. This was also observed from experiments where **SI Fig. S3 (a, b)** shows UV-visible spectra with two major absorption peaks and Tauc plot showing a bandgap of 2.8 eV in exfoliated $CrTe_3$. The enhanced piezo response was also due to flexoelectric effects as discussed in the above sections.

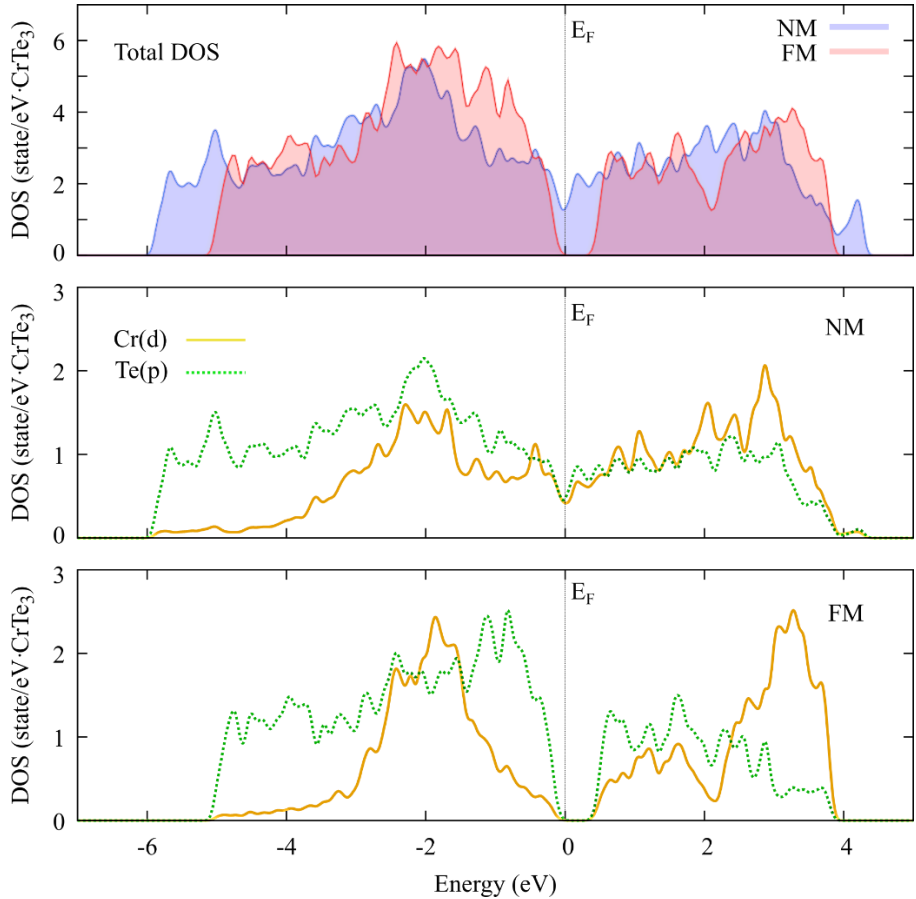

*Fig. 6: Bandgap calculations: Total and projected density of states for ferro- and non-magnetic states of $CrTe_3$ monolayer (for the sake of accuracy the SOC effects are included). Light green dashed line corresponds to the Cr d-orbitals, solid yellow line corresponds to the Te p-orbitals. The Femi energy $E_F$ is set to zero.*

Acoustic vibrations cause layer charge density effects in semiconducting 2D $CrTe_3$, resulting in a potential gradient in the opposite dielectric media (*Kapton*). This was observed from Zeta potential measurements, **Fig. 7a**. 2D $CrTe_3$ dispersion placed in Zeta cell was exposed to magnetic fields similar to that of MANG device operation (**Fig. 7a inset**). There was a decrease in potential, indicating more electronegative charges when exposed to magnetic fields than when they were not exposed. The potential with the highest electronegativity, -933 mV, was observed after 30 minutes of magnetic field exposure, as shown in **Fig. 7b**. This demonstrates that $CrTe_3$ flakes produce sufficient mutual repulsion for improved stability and more structured carriers in presence of magnetic field. For operation concerning the use of loudspeaker, the permanent magnets in the speaker polarizes the randomly oriented spins in a particular direction. The direction of orientation determines the magnetic anisotropy, thus resulting in a short-range FM ordering near room temperature. The magnetic field is important for spin alignment in ferromagnetic 2D $CrTe_3$ flakes. The aligned flakes provide more surface area and active sites for charges to flow through the MANG device. The device effects on electrochemical impedance of the device dielectric layer were studied with electrochemical impedance spectroscopic (EIS) measurements. The real (Z') and imaginary parts (Z'') were plotted for exposure of magnetic field effects as seen in **Fig. 7c**. The data is fitted with a simple circuit consisting of the series resistance ($R_s$), charge transfer resistance ($R_{ct}$) and capacitance ($Q$) (**Fig. 7c inset**). The calculated values from the fitted data are tabulated in **Table 1** (**SI section 2.5**). As soon as the device is exposed to magnetic field, we observe a sharp reduction in $R_{ct}$ and an increase in capacitance values to 254 MΩ and 603 pF, respectively. Whereas the $R_s$ remained in the range of 3.8 – 4 kΩ as observed from **Fig. 7d**.

The surface charge study shows that due to magnetic field, the charges were more stabilized and the layered $CrTe_3$ provided active sites for better charge accumulation (enhanced surface area) in the dielectric layer. The charge conduction was facilitated via the opposite dielectric layer, *Kapton* and through external Cu contacts. Hence making layered semiconductor ferromagnets such as 2D $CrTe_3$ potential magnetic-acoustic energy harvesters. The semiconductor ferromagnet has a strong coupling between the electron states of magnetic ions and electrons in semiconductor bands.

Magnetic ions can mediate between semiconductor bands (valance and conduction bands) in response to external magnetic fields [58]. The theoretical calculations show that the bandgap opening of ~ 1 eV has been observed for 2D CrTe$_3$ (**Fig. 6**). The flow of electrons in 2D CrTe$_3$ bands aids in charge conduction between bands in the material, while the dielectric media promotes in charge accumulation when exposed to permanent magnets (at 1000 Hz) from the loudspeaker.

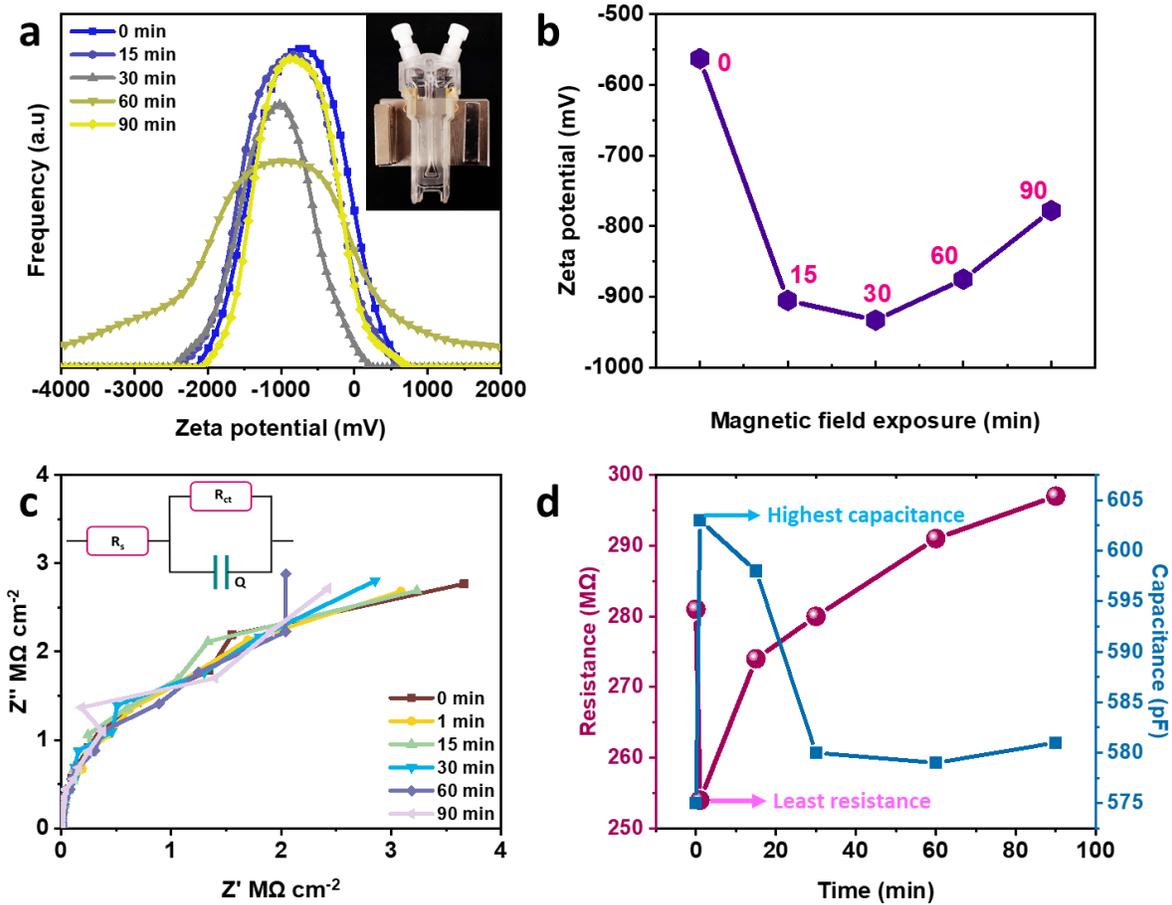

*Fig. 7: Surface charge interaction: (a) Zeta potential measurements; inset shows Zeta cell exposed to magnetic fields, (b) Plot of potential on exposure to magnetic fields, (c) Galvonostatic EIS of 2D MANG device; Circuit for fitting EIS data and (d) Charge transfer resistance ($R_{ct}$) and capacitance plots on exposure to magnetic fields.*

## 5.0 Conclusion

Asymmetric reduction of large-scale CrTe$_3$ flakes were produced by mechanical exfoliation as the van der Waals structure has a similar cleavage energy of 0.49 J m$^{-2}$ as that of graphene (0.5 J m$^{-2}$). The 2D CrTe$_3$ flakes with thickness range of ~4 nm were then magnetically characterized for its ferromagnetic nature with $T_c$ at ~224 K. The higher $T_c$ with high remnant magnetization is critical for device design operating around room temperature which was also confirmed from theoretical calculations. A magneto-acoustic nanogenerator (MANG) was fabricated which was able to harvest up to 6.4 V at low frequency range of 1000 Hz produced by a loudspeaker. The device showed maximum power density of 20 mW cm$^{-2}$ for operation at low-resonant frequency of 6.67 Hz which is higher than the existing low-frequency generators. The built in permanent magnets of the loudspeaker aligns the spins of magnetic ions, creating an exchange field that affects the spins of the band electrons (conduction and valence bands). Unidirectional anisotropy on the semiconducting layer was used to further modify the spin ordering in the device during operation at RT. The FM conditions were maintained for longer conditions via external magnetic fields slightly above critical temperature which was also observed from theoretical PDOS with SOC effects. Hence, the effect of the magnetic field on these helps generate charges in the dielectric layer. The charge facilitation was owed to the flexoelectric nature (strain gradient) enhancing the piezo-response in the 2D CrTe$_3$ during low-frequency operation. Therefore, the device can be used to potentially harvest magneto-acoustic waves emitted from a loudspeaker and other vibrating devices operating at low frequencies.